\pdfoutput=1

\documentclass[11pt]{article}

\usepackage[]{EMNLP2023}

\usepackage{times}
\usepackage{latexsym}
\usepackage{tikz}

\usepackage[T1]{fontenc}

\usepackage[utf8]{inputenc}
\usepackage{listings}
\usepackage{microtype}
\usepackage{booktabs}
\usepackage{tabularx}
\usepackage{amssymb}
\usepackage[most]{tcolorbox}

\usepackage{graphicx}   
\newcommand{\chatlogo}{\raisebox{-.4\height}{\includegraphics[height=3.5ex]{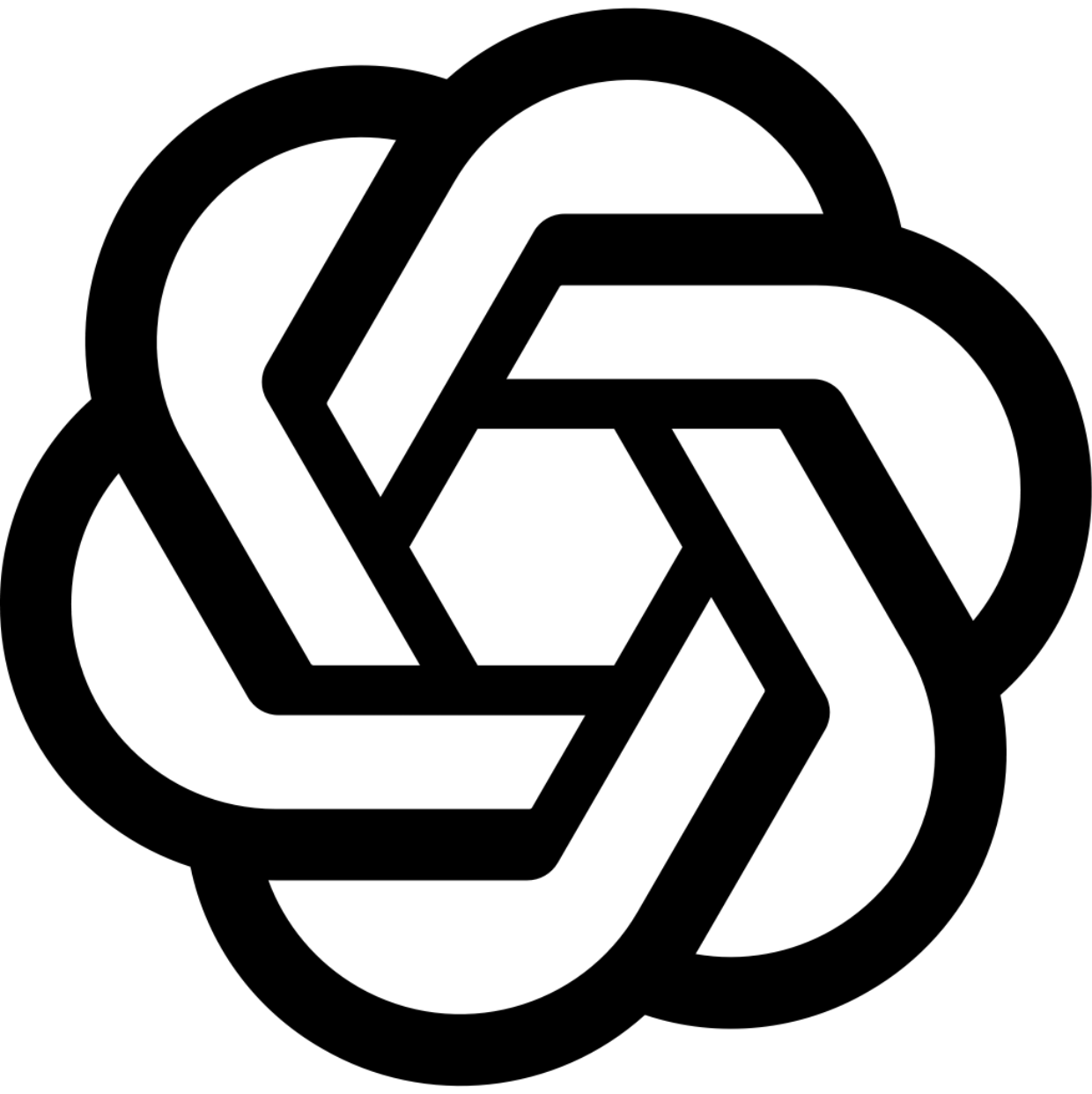}}}

\usepackage{pifont}
\usepackage[]{xcolor}

\usepackage{multirow} 
\usepackage{amsmath}
\usepackage{array}
\newcolumntype{C}[1]{>{\centering\arraybackslash}p{#1}}

\usepackage{inconsolata}

\title{StockSim: A Dual-Mode Order-Level Simulator for Evaluating Multi-Agent LLMs in Financial Markets}

\author{
    Charidimos Papadakis, Giorgos Filandrianos, Angeliki Dimitriou, \\
    \textbf{Maria Lymperaiou, Konstantinos Thomas, Giorgos Stamou}\\ 
    School of Electrical and Computer Engineering, AILS Laboratory\\
    National Technical University of Athens \\
    \texttt{\href{mailto:harrypapadakis02@gmail.com}{harrypapadakis02@gmail.com}}, \\ 
    \texttt{\{\href{mailto:geofila@ails.ece.ntua.gr}{geofila},\href{mailto:angelikidim@ails.ece.ntua.gr}{angelikidim}, \href{mailto:marialymp@ails.ece.ntua.gr}{marialymp}, \href{mailto:kthomas@ails.ece.ntua.gr}{kthomas}}\}@ails.ece.ntua.gr,
    \\
    \texttt{\href{mailto:gstam@cs.ntua.gr}{gstam@cs.ntua.gr}}\\
}
\definecolor{ForestGreen}{RGB}{34,139,34}
\definecolor{BrickRed}{rgb}{0.8, 0.25, 0.33}
\newcommand{\cmark}{\textcolor{ForestGreen}{\ding{51}}}               
\newcommand{\xmark}{\textcolor{BrickRed}{\ding{55}}}               
\begin{document}
\maketitle

\begin{abstract}
We present \textbf{\textsc{StockSim}}, an open-source simulation platform for systematic evaluation of large language models (LLMs) in realistic financial decision-making scenarios. Unlike previous toolkits that offer limited scope, \textbf{\textsc{StockSim}} delivers a comprehensive system that fully models market dynamics and supports diverse simulation modes of varying granularity. It incorporates critical real-world factors, such as latency, slippage, and order-book microstructure, that were previously neglected, enabling more faithful and insightful assessment of LLM-based trading agents.
An extensible, role-based agent framework supports heterogeneous trading strategies and multi-agent coordination, making \textsc{StockSim} a uniquely capable testbed for NLP research on reasoning under uncertainty and sequential decision-making. We open-source all our code at \url{https://github.com/harrypapa2002/StockSim}.
\end{abstract}

\section{Introduction}

Financial markets present complex, dynamic environments characterized by high uncertainty, consequential decisions, and measurable outcomes \cite{yadav2020measuring, rudkin2023uncertainty, nafiu2025risk}. As large language models (LLMs) have demonstrated significant proficiency in sequential reasoning and decision-making tasks \cite{chen-etal-2024-efficient, liu2025fin}, systematically evaluating these models within realistic financial scenarios has emerged as a crucial research direction for NLP.

However, the NLP community currently faces substantial obstacles due to a lack of standardized and openly accessible platforms specifically designed for rigorous evaluation of LLMs in realistic trading contexts \cite{li2024investorbench, lu2025bizfinbench}. Common evaluation practices that rely on static benchmark datasets inadvertently risk data leakage, as these datasets or similar financial texts often appear in LLM training corpora \cite{dong-etal-2024-generalization, singh2024evaluation, white2024livebench}. Consequently, performance metrics become inflated, and the models fail to generalize effectively to genuinely unseen scenarios, creating unrealistic expectations and potential financial risks when deployed.

Existing evaluation platforms further compound these limitations, as highlighted in Table \ref{tab:framework_feature_comparison}. Frameworks such as Backtrader\footnote{\url{https://pypi.org/project/backtrader/}} and FinRL \cite{liu2020finrl, liu2022finrlmeta} offer extensive historical backtesting but abstract away crucial trading microstructure aspects like latency and detailed order-book dynamics. Conversely, platforms like ABIDES \cite{byrd2020abides}, PyMarketSim \cite{mascioli2024financial}, and JAX-LOB \cite{frey2023jax} simulate precise order-level market mechanics but depend heavily on expensive, limited, tick-level datasets, restricting their practical applicability and scalability. Additionally, frameworks designed for multi-agent LLM coordination, such as TradingAgents \cite{xiao2024tradingagents}, often use highly simplified market representations based solely on coarse historical data, omitting realistic execution and latency considerations critical for evaluating LLM behavior.

\begin{table*}[t!]
\vskip -0.08in
\centering
\small
\setlength{\tabcolsep}{4pt}
{%

\begin{tabular}{lC{1.8cm}C{1.3cm}C{1.3cm}C{1.3cm}C{1.3cm}C{1.5cm}C{1.3cm}C{1.5cm}}
\hline
\shortstack{\textbf{Framework}} &
\shortstack{\textbf{Execution} \\ \textbf{Granularity}} &
\shortstack{\textbf{Async} \\ \textbf{Latency}} &
\shortstack{\textbf{Real‐time} \\ \textbf{LOB}} &
\shortstack{\textbf{History} \\ \textbf{Back‐test}} &
\shortstack{\textbf{No-code} \\ \textbf{Setup}} &
\shortstack{\textbf{LLM Agent} \\ \textbf{Support}} &
\shortstack{\textbf{External} \\ \textbf{News}} &
\shortstack{\textbf{Multi} \\ \textbf{Instrument}} \\

\hline
\textbf{StockSim (ours)} & Order & \cmark & \cmark & \cmark & \cmark & \cmark & \cmark & \cmark\\
ABIDES            & Order & $\sim$ & \cmark & \cmark & \xmark & \xmark & \xmark & \cmark\\
PyMarketSim       & Order & \xmark & \cmark & \xmark & \xmark & \xmark & \xmark & \cmark\\
JAX‐LOB           & Order & \xmark & \cmark & \cmark & \xmark & \xmark & \xmark & \cmark\\
FinRL / Meta      & Bar   & \xmark & \xmark & \cmark &  $\sim$ & \xmark & $\sim$ & \cmark\\
TradingAgents     & Daily & \xmark & \xmark & \cmark & \xmark & \cmark & \cmark & \cmark\\
\hline
\end{tabular}%
}
\caption{
Feature comparison of open-source trading simulators.
\textit{Legend.}  
\cmark: supported;  
\xmark: not supported;  
$\sim$: partial or approximate support. }

\label{tab:framework_feature_comparison}
\end{table*}

The fragmented landscape forces researchers either to simplify market interactions unrealistically or to invest significant resources developing custom, often proprietary pipelines, which hinder reproducibility, fair comparisons, and collective research progress in NLP-driven financial decision-making.
To overcome such challenges, we introduce \textbf{\textsc{StockSim}}, a unified, open-source, fully pre-configured  platform explicitly designed to rigorously assess LLM behavior in realistic, dynamic financial scenarios. It integrates two complementary simulation modes behind a single interface: 
(i) an \textbf{order-level }execution mode that simulates fine-grained market behavior capturing latency, queue dynamics, and microstructural dynamics; and (ii) a \textbf{candlestick-level} (price-bar-level) execution mode that enables scalable evaluation while abstracting away low-level market effects.

\textsc{StockSim} shifts the research focus from evaluation infrastructure development to core NLP-driven agent design and experimentation. Specifically, it enables: 1) Comprehensive ordel-level simulation, capturing essential trading dynamics from detailed tick-level events to aggregated price bars. 2) Flexible, high-throughput bar-level execution across diverse market scenarios, assets (stocks, cryptocurrencies), and temporal resolutions. 3) Agent utilization of external multi-modal information, such as news sentiment and financial reports, facilitating realistic NLP experimentation.
4) Robust, production-grade infrastructure providing authentic market data integration, real-time technical indicator calculations, and detailed agent performance tracking.




\section{Background}

\textsc{StockSim} exposes agents to realistic trading dynamics, including order execution via a limit-order book (LOB), delays from latency, price shifts from market impact, and costs like slippage. Strategies rely on historical OHLCV (candlestick) data, derived technical indicators, and real-time market mechanics - making familiarity with these concepts essential for interpreting agent behavior.

\textbf{Order types.}
A \emph{market order} executes immediately against the best prices available;  
a \emph{limit order} is queued and only fills at its stated price (limit) or better. Typically used to enter trades;  
a \emph{stop order} is executed as a market order once a trigger price is hit, typically used to exit trades.

\textbf{Limit-order book (LOB).} 
The LOB is the continuously updated queue of pending buy (bid) and sell (ask) limit orders at each price level.  It is responsible for storing these orders and facilitating their resolution when matching conditions are met. It drives price discovery and is the core of \textsc{StockSim}’s real-time simulator.

\textbf{Latency \& market impact.}
\emph{Latency} is the delay between submitting an order and that order reaching the order-matching system; even sub-second differences matter in  modern electronic environments.  
\emph{Market impact} is the price movement an order itself induces. Large market orders may clear out substantial amounts of pending orders in the limit-order book, moving the asset's price.

\textbf{Slippage.}
The difference between the prices at which an order is submitted by the trader (or trading system) and the price it actually gets fulfilled, is called \emph{slippage}; it is the consequence of latency and market impact effects and is a major source of hidden and unpredicted trading costs.

\textbf{Market microstructure.}
The term covers statistical regularities of order placements, cancellations, spread dynamics, queue imbalance, and how they interact to form price.
\textsc{StockSim}’s order-book engine reproduces these dynamics so that agents must cope with queue position, partial fills, and other microstructure realities.

\textbf{OHLCV (candlestick) bars \& timeframes.}
Historical data are often stored as \emph{open, high, low, close, volume} (OHLCV) tuples, commonly called \emph{candlesticks} due to their appearance on a chart: (i) \textbf{Open} – the first traded price in the bar; (ii) \textbf{High} – the maximum traded price; (iii) \textbf{Low} – the minimum traded price; (iv) \textbf{Close} – the last traded price; 
(v) \textbf{Volume} – total quantity traded during the bar.
The bars have a constant start/end time (i.e. 5 minutes, 1 day, etc.) referred to as the chart's \emph{timeframe}. \textsc{StockSim}’s candlestick engine replays any asset at resolutions from one minute to one day (and beyond).


\textbf{Technical indicators.}
Deterministic functions of past price or volume, such as moving averages, Relative Strength Index (RSI) or Average True Range (ATR) that serve as numeric features of concentrated information, used heavily in 
trading. More details can be found in Appendix~\ref{app:indicators}. 

\begin{figure*}[t!]
\vskip -0.12in
    \centering    \includegraphics[width=0.82\textwidth]{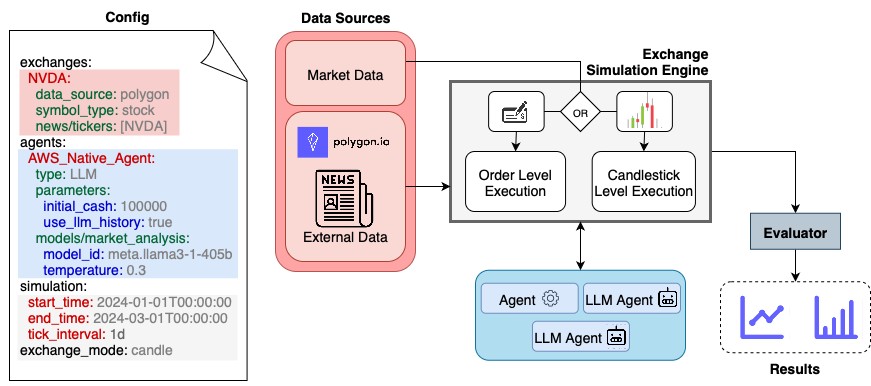}
    \caption{Overview of \textsc{StockSim}'s system architecture and input/output scheme. Modules are color-coded by function and mapped to corresponding blocks in the centralized config file. This design supports flexible, code-free customization of simulation parameters, agent behavior, and data sources.}
    \label{fig:architecture}
\end{figure*}

\section{System Architecture}


\textsc{StockSim} employs a modular, asynchronous architecture designed around four core components that enable comprehensive LLM evaluation  in realistic trading environments. Figure \ref{fig:architecture} illustrates the system's data flow and component interactions, highlighting two execution mechanisms—\emph{order level} and \emph{candlestick level execution} —seamlessly integrated with shared modules for market data retrieval, indicator computation, news/fundamentals integration, and agent interactions. This design ensures consistency, flexibility, and scalability, supporting diverse experimental setups and facilitating reproducible experimentation on sequential decision-making in financial contexts.

\subsection{Exchange Simulation Engine}






The core component of \textsc{StockSim} is the \textit{Exchange Simulation Engine}, which asynchronously manages and coordinates the simulated trading environment. Its primary responsibilities include: (1) receiving and processing agent actions (e.g., order placements); (2) simulating realistic market dynamics for order execution; (3) computing and disseminating relevant market indicators; and (4) providing agents with timely access to market and external information (e.g. news, corporate events).

The \textit{Engine} acts as the central intermediary between data sources and trading agents. It does not directly store external data; in contrast, it routes the data dynamically from their respective sources to agents upon request, actively maintaining internal states related to orders and trades, including execution status and market impact.
Each agent runs as a separate process and communicates asynchronously with the Engine to submit orders, request data, or receive market updates. This asynchronous communication is managed via RabbitMQ, an advanced message broker that ensures reliable message delivery and scalable communication\footnote{\url{https://www.rabbitmq.com/}}.
Moreover, the \textit{Engine} supports two distinct ways of resolving orders submitted by agents, designed to accommodate various research scenarios:

\textbf{Order Level Execution}
emulates real market behavior by operating directly on the LOB, where the agent submits limit or market orders that interact with a stream of order book events (placements, cancellations, executions). Orders are matched based on price-time priority: e.g., a buy limit order at \$100 will execute only if a sell order exists at \$100 or lower; otherwise, it queues until matched or canceled. Execution may be full or partial, depending on available volume. The environment updates tick-by-tick, capturing fine-grained dynamics such as queue position, order interleaving with other market participants, and the impact of latency between action submission and book update. This level offers high realism and is critical for evaluating strategies sensitive to microstructure effects.

\textbf{Candlestick Level Execution}
places orders based on aggregated candlestick data (OHLCV). That is, if the agent submits an order at a price that falls within the range of a given candle, the order can be executed; otherwise, it cannot. We adopt this approach as it provides access to a larger dataset, enabling testing over longer historical periods. Moreover, most LLMs are evaluated under this setting \cite{li-etal-2024-cryptotrade}. Despite being widely used, mainly due to its simplicity, this mode fails to capture critical dynamics, such as latency and other microstructural elements of real markets.

Both execution mechanisms consistently provide agents with computed market indicators 
(e.g., SMA, EMA, RSI, VWAP) derived from real-time or historical market data, enhancing 
agents’ decision-making capabilities (see Appendix~\ref{app:indicators}).

\subsection{Data Sources}




\textsc{StockSim} distinguishes between two primary categories of data: 
(1) \textbf{market data}, which include price, volume, and order-flow information; and 
(2) \textbf{external data}, such as news, corporate actions, and fundamental metrics.  
The \textit{Exchange Simulation Engine} orchestrates these inputs asynchronously, delivering them to agents in simulation time.

\paragraph{Market Data.}
\textsc{StockSim} supports two types of market data: detailed \textit{order-level} data and simplified \textit{bar-level} (candlestick) data.

In the \emph{candlestick level execution}, the data is provided as aggregated summaries i.e. OHLCV bars, obtained from general data sources like Alpha Vantage and Polygon.io\footnote{\url{https://www.alphavantage.co};\;\url{https://polygon.io}}. Because these summaries do not include detailed, within-bar price movements, \textsc{StockSim} simulates realistic price paths within each bar. This allows agents to place conditional orders (like stop losses) that execute plausibly, even though exact moment-to-moment data is not available.
In the \emph{order level execution}, each market action, such as placing, changing, or cancelling an order is individually tracked. These detailed events come either from datasets like LOBSTER\footnote{\url{https://lobsterdata.com}} or from logs created during the simulation. Each event has precise timestamps (milliseconds), allowing realistic simulation of
latency
and 
slippage.

\paragraph{External Data.}
Agents may request news headlines, earnings calendars, splits, dividends, or fundamental ratios at any simulation step.  
These streams are supplied through the same provider set (Alpha Vantage, Polygon, etc.) and exposed via a unified query interface implemented by the Exchange Simulation Engine.  
This abstraction lets agents reason over time-sensitive, multi-modal inputs and supports the development of more interpretable, information-driven trading strategies.  
Because each provider is wrapped by a lightweight adapter that maps its payloads to \textsc{StockSim}’s canonical schema, adding a new API is as simple as contributing a single Python file, ensuring the platform can evolve alongside the data ecosystem.

\subsection{Agent}





Agents are the research object in \textsc{StockSim}.  Regardless of which \emph{execution
engine} mode is plugged in, every agent interacts with the simulator through the
same asynchronous message API; only the \emph{engine} decides how an order is
ultimately filled.  This separation allows a single agent implementation - 
written once in Python - to be stress-tested on both the order level and
candlestick level execution mode without code changes.

\paragraph{Core Capabilities.}
Each agent may:

1) \textbf{Subscribe to data streams.}
Agents request snapshots or streaming updates of
\emph{(i)} market state (order book depth or OHLCV bars),
\emph{(ii)} technical indicators produced on-the-fly, and
\emph{(iii)} external content such as news, corporate events etc. 

2) \textbf{Submit and cancel orders.}
The message schema supports \texttt{MARKET}, \texttt{LIMIT}, and
\texttt{STOP} instructions of arbitrary size.
 In the \textbf{Order Level Execution} mode the order is routed
to a price–time priority matcher and may experience queueing,
latency, and market impact.  In the \textbf{Candlestick Level Execution} mode the same message is interpreted by
by whether the price action within a candle crosses the pending order price.

3) \textbf{Receive execution outcomes and portfolio updates.}
Agents immediately receive feedback about their submitted orders, including confirmations of successful trades (fills), rejections if an order could not be executed, cancellations if the agent withdraws an order, and updates about their profit-and-loss (P\&L). This ensures agents have timely information to adapt their decisions.

4) \textbf{Log reasoning.}
Optional free-form “explanation’’ strings can accompany every order;
these are preserved by the engine and can be inspected offline to
analyse LLM rationale and decision trace.

\paragraph{Multi-Agent and Specialist Roles.}
\textsc{StockSim} includes a modular \texttt{LLMTradingAgent} that delegates decision-making to a team of specialist LLMs (e.g., market-technical analyst, news analyst, fundamental analyst). Each analyst operates with its own prompt template, memory context, and reasoning function. While adding new analyst roles does require lightweight code changes, such as implementing a new analyst class and registering it in the coordinator, the process is intentionally simple, well-documented, and configuration-driven. The modular structure ensures that agent internals remain decoupled from the simulation engine, making it easy to test new multi-agent frameworks or LLM coordination setups with minimal friction.

This design enables researchers to:
\emph{(i)} rapidly prototype new agent structures, including macroeconomic, sentiment, or reflection-based analysts;
\emph{(ii)} experiment with different backbones or prompting techniques per role;
\emph{(iii)} plug analysts into different coordination strategies, such as voting, tree-of-thought, or chain-of-experts;
\emph{(iv)} conduct clean ablation studies by toggling analysts or modifying their configuration in one place.

\paragraph{Research Convenience.}
\textsc{StockSim} maintains a unified interface across  simulation engines: switching between order level and candlestick level execution
requires only a configuration change, not architectural rewrites, isolating research focus on core questions like prompt engineering,  reasoning, or analyst collaboration, without being burdened by low-level simulation mechanics. 
In technical terms, we implement a base class per agent, enabling easy adaptation to specific  needs. We also provide pre-configured wrappers for widely used LLMs, including LLaMA, OpenAI’s offerings (e.g., GPT-o4, o3), and Anthropic’s models (e.g., Claude Sonnet, Haiku, Opus).

\subsection{Evaluator}


The \textit{Evaluator} component subscribes to all trade executions, recording a complete history of positions, cash and realized P\&L (Profit \& Loss).  Once the simulation finishes, it computes a concise set of core performance metrics, such as overall return, risk‐adjusted ratios (e.g.\ Sharpe), drawdown, and basic trade statistics, finally packaging them into a uniform report. A list of predefined metrics and their definitions can be found in Appendix~\ref{app:Metrics}.
Crucially, the metric evaluation system is designed to be \textbf{fully extensible}. Users can seamlessly integrate custom performance measures, such as tail-risk, turnover, or regime-specific statistics, by registering additional evaluation components, all without modifying the core simulation engine.

\begin{figure}[t!]
\vskip -0.08in
  \centering
  \includegraphics[width=1.02\linewidth]{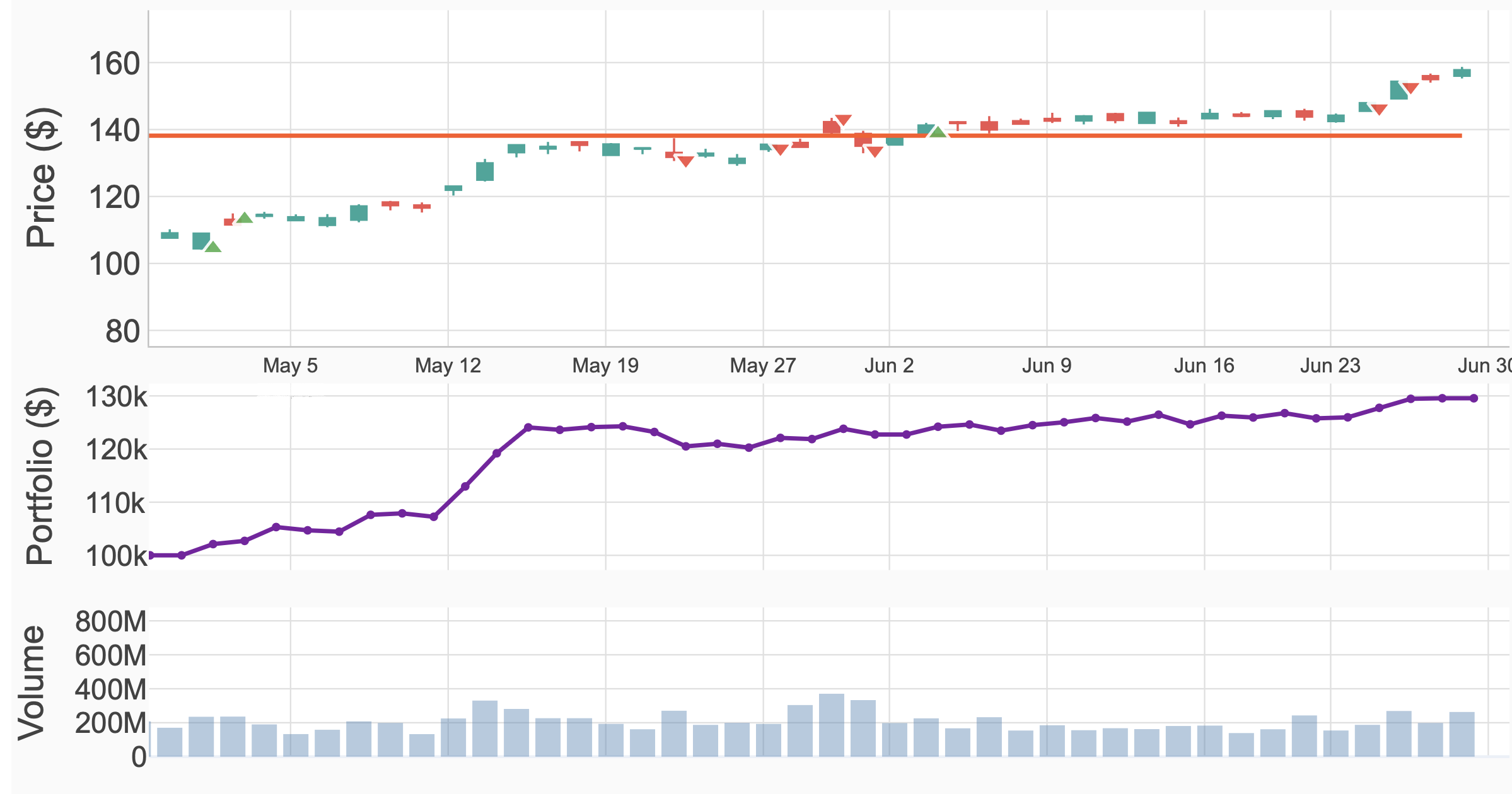}
\caption{Performance of GPT-o3, showing executed trades and portfolio value evolution. 
Sell actions are marked with {\color{BrickRed}$\blacktriangledown$}, while buy actions are marked with {\color{ForestGreen}$\blacktriangle$}.}
  \label{fig:nvda-performance}
\end{figure}

For visual diagnostics, the \textit{Evaluator} provides several useful outputs, such as equity curves showing portfolio value changes over time, candlestick charts highlighting executed trade entries and exits clearly marked on the price data, and comprehensive summary tables of key trading performance metrics.
These outputs can be directly generated by \textsc{StockSim}'s built-in plotting utilities or exported in JSON format for further analysis. An example of these diagnostics for a trading session involving NVIDIA (NVDA) using the o3 model evaluated at the candlestick level is shown in Figure~\ref{fig:nvda-performance}. Additionally, all  figures provided by \textsc{StockSim} offer interactive features (zoom-in/out and hover-to-display details like OHLCV data) and model explanations, advancing interpretability of decision-making. Examples can be found in Appendix~\ref{app:visualition}.

\section {Evaluation}
\textbf{Scalability and Consistency}
of \textsc{StockSim} are evaluated through a series of controlled simulation tests using varying numbers of \textit{deterministic} agents.
Each agent follows predefined strategies, such as moving average crossovers or buy-and-hold, allowing us to observe the simulation engine’s behavior under repeatable conditions. To ensure that the evaluation reflects only the core behavior of the engine, we exclude LLMs, which introduce variability in latency, resource usage, and output consistency due to differences in deployment mode, reasoning strategy, and stochastic outputs.
\begin{figure}[t!]
\vskip -0.12in
  \centering
  \includegraphics[width=1\linewidth]{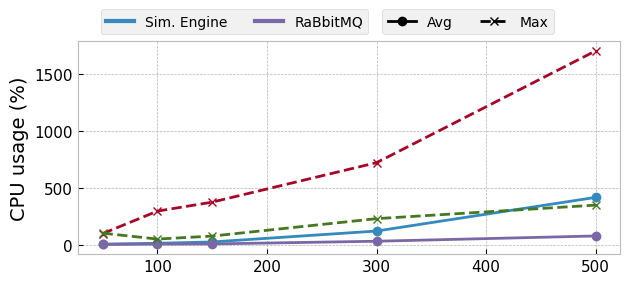}
  \includegraphics[width=1\linewidth]{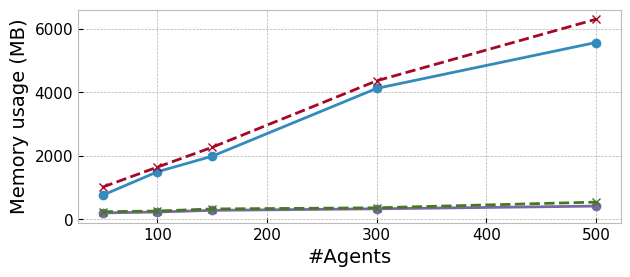}
\caption{System performance metrics (memory/CPU usage) for varying numbers of deterministic agents.}
  \label{fig:performance-metrics}
\end{figure}
The results confirm \textsc{StockSim}’s consistency: across all runs, simulation outputs (including order placements, executions, and performance metrics) \textbf{remain identical}. This repeatability empirically verifies the platform’s deterministic behavior and validates its correctness, since any deviation would indicate flaws in the design or execution logic.

Scalability is assessed by monitoring system-level metrics during each run, including CPU utilization across all cores and memory usage (in MB) for both the simulation engine and RabbitMQ. Results across agent configurations are presented in Figure \ref{fig:performance-metrics}, confirming that \textsc{StockSim} \textbf{scales almost linearly} up to $\sim$150 agents: the simulation container’s mean CPU load increases from 8\% to 27\%, while memory usage rises from 0.8 GB to 2 GB, both roughly proportional to the agent count. Beyond this point, the workload becomes super-linear: at 300 and 500 agents, mean CPU usage surges to 123\% and 418\%, and memory climbs to 4.1 GB and 5.6 GB, respectively, with peak values reaching nearly four times the averages.

Despite this growth, the resource demands of the simulation framework remain modest; even at maximum load, usage peaks at 5.6 GB of RAM and a few CPU cores. All experiments are conducted on a MacBook Pro with an Apple M3 Pro chip (11-core CPU) and 18 GB of unified memory, underscoring \textsc{StockSim}’s efficiency. Running 500 concurrent LLM agents in parallel is practically infeasible on such hardware, whereas this analysis demonstrates that \textsc{StockSim} can handle such scale with ease.

\begin{table}[t!]
\vskip -0.08in
  \centering
  \small
  \begin{tabular}{l|c c}
    \hline
    \textbf{Metric}                & \textbf{\chatlogo{} o4-mini} & \textbf{\chatlogo{} o3} \\
    \hline
    ROI  ($\uparrow$)                          & 0.0734            & 0.2956            \\
    Sharpe Ratio - SR ($\uparrow$)                   & 0.1652            & 0.376             \\
    Annualized SR ($\uparrow$)       & 2.6218            & 5.9682            \\
    Sortino Ratio ($\uparrow$)                 & 0.2868            & 1.0587            \\
    Win Rate   ($\uparrow$)                    & 0.6667            & 1.0               \\
    Profit Factor ($\uparrow$)                 & 2.3691            & 999.0             \\
    Max Drawdown ($\downarrow$)                & 0.0306            & 0.0323            \\
    Num Trades                                 & 31                & 9                 \\
    Num Closed Trades                          & 21                & 6                 \\
    Total Traded Volume                        & 931,416.775       & 368,306.25        \\
    Average Trade Size                         & 30,045.70         & 40,922.92         \\
    ROIC                                       & 0.0151            & 0.1633            \\
    Profit per Trade  ($\uparrow$)             & 258.47            & 4,520.13          \\
    Last Portfolio Value   ($\uparrow$)        & 107,338.30        & 129,556.75        \\
    Realized P\&L                       & 5,427.80          & 27,120.75         \\
    \hline
  \end{tabular}
  \caption{Summary of trading performance metrics (Appendix \ref{app:Metrics}) for GPT-o4-mini and GPT-o3.}
  \label{tab:nvda-metrics}
\end{table}

\paragraph{LLM Trading Behavior}

To demonstrate the ease with which insights about model behavior can be extracted using \textsc{StockSim}, we run a simulation for two LLMs, GPT-o4-mini and GPT-o3, using the same prompt (Appendix \ref{app:prompts}) on the NVIDIA stock over a two-month period, from April 28, 2025, to June 28, 2025. The simulation assumes daily trading, with orders placed before market open. The  results based on the performance metrics provided by \textsc{StockSim}, are presented in Table~\ref{tab:nvda-metrics}, revealing distinct trading patterns and strategic behaviors between LLMs. Metrics such as ROI, Profit per Trade, and Profit Factor highlight that GPT-o3 pursues a more selective trading strategy characterized by fewer, larger-sized positions with higher conviction, resulting in greater profitability and reduced downside risk, as demonstrated by its superior Sortino Ratio and perfect Win Rate. Conversely, GPT-o4-mini exhibits a more active trading style,  evidenced by its higher number of trades and greater traded volume, indicating frequent market interactions but lower profit efficiency per transaction. The contrasting Sharpe Ratio and Annualized Sharpe Ratio further underscore GPT-o3’s superior ability to maintain consistent, risk-adjusted returns over time, while GPT-o4-mini’s lower metrics suggest that its strategy involves more frequent but less decisive market positions. Overall, the \textit{evaluator}'s results  effectively capture and distinguish the underlying strategic differences between the two LLMs, allowing clear interpretation of their respective trading behaviors. Importantly, we are able to obtain these results\textbf{ without writing any code}, paving the way for exploring more LLM-driven trading strategies.

\section{Conclusion}

\textsc{StockSim} represents a significant advancement in NLP research infrastructure, providing a sophisticated platform for studying LLM abilities in realistic, multi-agent, temporal reasoning scenarios. By combining production-grade financial simulation with comprehensive NLP evaluation tools, \textsc{StockSim} enables research that bridges research experiments with real-world deployment requirements. The open-source availability and extensive documentation ensure broad accessibility for advancing our understanding of LLM behavior in complex, consequential decision-making environments.

\section*{Limitations}

While \textsc{StockSim} provides a comprehensive framework for LLM evaluation in financial domains, some limitations should be noted. The platform requires substantial computational resources for multi-agent simulations and may have scalability constraints for very large agent populations. Market simulation, while realistic, cannot fully capture all complexities of actual trading environments including liquidity constraints and market impact. The current evaluation metrics, while comprehensive, may not capture all aspects of decision quality relevant to financial applications. Additionally, the platform's focus on financial markets may limit generalizability to other sequential decision-making domains.

\section*{Ethics Statement}

\textsc{StockSim} is designed for research purposes and uses simulation environments that do not interact with real financial markets, eliminating concerns about market manipulation. The platform promotes responsible AI development by providing tools to systematically study LLM reliability and consistency, as well as model explanations that can be further assessed by humans for their credibility and accuracy. All market data is obtained through legitimate commercial APIs with appropriate licensing. The open-source nature of the platform ensures transparency and enables broader scrutiny of the evaluation methodologies employed.

\section*{Acknowledgements}

We thank the contributors to the open-source libraries that made this work possible, including the maintainers of RabbitMQ, asyncio, and various LLM APIs. We also acknowledge Polygon.io for providing comprehensive market data access that enables realistic financial simulations. This work was supported by the Hellenic Foundation for Research and Innovation (HFRI)  under the 5th Call for HFRI PhD Fellowships
(Fellowship Number 19268).
\bibliography{emnlp2023}
\bibliographystyle{acl_natbib}

\appendix
\section{Market Indicators in \textsc{StockSim}} \label{app:indicators}

Agents in \textsc{StockSim}, including both benchmark and LLM agents, receive a compact \emph{market-embedding} consisting of hand-crafted scalar indicators summarizing critical aspects of the latent market state. These indicators serve as low-entropy numeric tokens, simplifying the sequential-reasoning challenge for systematic evaluation across diverse agent types.
The embedding includes the following categories:

\begin{itemize}

\item \textbf{Trend Indicators:} Capture directional movement in market prices, allowing agents to infer the prevailing trend.
\begin{itemize}
\item \emph{Simple Moving Average (SMA)}: Average price over a specified look-back window, smoothing out short-term fluctuations.
\item \emph{Exponential Moving Average (EMA)}: Weighted average emphasizing recent prices, responsive to short-term trend changes.
\end{itemize}

\item \textbf{Momentum Indicators:} Measure the strength and velocity of price movements, enabling agents to recognize acceleration or deceleration in trends.
\begin{itemize}
\item \emph{Relative Strength Index (RSI)}: Quantifies momentum by comparing average gains and losses, identifying overbought or oversold conditions.
\item \emph{Moving Average Convergence Divergence (MACD)}: Highlights momentum shifts by comparing short-term and long-term EMAs.
\end{itemize}

\item \textbf{Volatility Indicators:} Reflect market uncertainty or risk by measuring price variability.
\begin{itemize}
\item \emph{Average True Range (ATR)}: Computes volatility based on price ranges, useful for assessing market turbulence.
\item \emph{Bollinger Bands}: Envelopes around SMA, indicating volatility expansion or contraction.
\end{itemize}

\item \textbf{Volume and Micro-Structure Indicators:} Capture trading activity intensity and structural nuances of market participation.
\begin{itemize}
\item \emph{Volume Weighted Average Price (VWAP)}: Reflects the average traded price weighted by volume, indicating liquidity-driven price levels.
\item \emph{Order Book Imbalance}: Measures differences in bid and ask quantities, signaling buying or selling pressure.
\end{itemize}

\item \textbf{Support-Resistance Indicators:} Identify critical price levels at which market dynamics historically reverse or accelerate.
\begin{itemize}
\item \emph{Historical Support Levels}: Previous price levels where buying activity halted declines.
\item \emph{Historical Resistance Levels}: Prior price ceilings where selling activity halted rallies.
\end{itemize}

\end{itemize}

By encapsulating rich market dynamics into concise numeric tokens, \textsc{StockSim}'s market-embedding reduces complexity, facilitating effective sequential decision-making research.

\section{Trading Performance Metrics}
\label{app:Metrics}

The definitions of the predefined metrics used by StockSim for evaluation are as follows\footnote{\href{https://www.investopedia.com}{https://www.investopedia.com}}:

\begin{itemize}
    \item \textbf{ROI (Return on Investment):} ROI measures the profitability of an investment relative to its cost. It is calculated as the ratio of net profit to the initial capital investment, indicating the efficiency of the strategy.

    \item \textbf{Sharpe Ratio (SR):} The Sharpe Ratio evaluates the risk-adjusted return of an investment. It is defined as the ratio of the excess return over the risk-free rate to the standard deviation of returns. A higher Sharpe Ratio suggests a more favorable risk-return profile.

    \item \textbf{Annualized Sharpe Ratio (Annualized SR):} This metric adjusts the Sharpe Ratio to an annual scale, enabling comparison across different timeframes and investment durations.

    \item \textbf{Sortino Ratio:} The Sortino Ratio refines the Sharpe Ratio by considering only the downside deviation (negative volatility), thereby focusing on harmful risk. It is the ratio of excess return to the standard deviation of negative returns.

    \item \textbf{Win Rate:} The Win Rate represents the proportion of profitable trades out of the total number of executed trades. It reflects the consistency and reliability of the strategy.

    \item \textbf{Profit Factor:} The Profit Factor is defined as the ratio of gross profits to gross losses across all trades. Values greater than 1 indicate a profitable strategy; higher values imply greater efficiency in managing risk and reward.

    \item \textbf{Max Drawdown:} Maximum Drawdown measures the largest peak-to-trough decline in portfolio value during the evaluation period. It serves as an indicator of downside risk and potential capital loss.

    \item \textbf{Number of Trades:} This is the total number of executed trades, including both open and closed positions.

    \item \textbf{Number of Closed Trades:} This refers to the total count of trades that have been fully executed and settled within the evaluation period.

    \item \textbf{Total Traded Volume:} The aggregate monetary value of all executed trades. This metric reflects the scale and activity of the trading strategy.

    \item \textbf{Average Trade Size:} The mean monetary value of executed trades. It provides insight into the average scale of trade operations.

    \item \textbf{ROIC (Return on Invested Capital):} ROIC quantifies the return generated on the capital that was actually deployed in trades. It provides a precise view of the strategy’s capital efficiency.

    \item \textbf{Profit per Trade:} The average net profit generated per closed trade. This metric highlights the effectiveness of individual trade decisions.

    \item \textbf{Last Portfolio Value:} The total value of the portfolio at the conclusion of the evaluation period. It reflects the cumulative financial outcome of the strategy.

    \item \textbf{Realized P\&L:} The net profit or loss from closed trades. This metric does not include unrealized gains or losses from open positions.
\end{itemize}

\section{Technical Indicators}
\label{app:indicators}
Technical indicators are deterministic mathematical functions derived from historical price and/or volume data \cite{wilder450new, Cartea2015RISK}. They transform raw market data into condensed numeric features that aim to reveal trends, momentum, volatility, or potential reversals. These indicators are widely used by both human traders and algorithmic systems to inform trading decisions and strategy development. Indicators can be easily extended in the StockSim framework; some predefined ones are presented below.

\paragraph{Moving Averages (MA).}
A moving average smooths out price data by calculating the average of past prices over a fixed window. It helps identify trend direction and reduce short-term noise.
\begin{itemize}
    \item \textbf{Simple Moving Average (SMA):}
    \begin{equation}
    \text{SMA}_n = \frac{1}{n} \sum_{i=1}^{n} P_i
    \end{equation}
    where $P_i$ is the closing price at time step $i$ and $n$ is the number of periods.

    \item \textbf{Exponential Moving Average (EMA):}
    A weighted average that gives more importance to recent prices, making it more responsive to recent changes. The weighting decreases exponentially for older prices.
\end{itemize}

\paragraph{Relative Strength Index (RSI).}
RSI is a momentum oscillator that measures the speed and magnitude of recent price changes to evaluate overbought or oversold conditions. RSI values range from 0 to 100.
\begin{equation}
    RSI = 100 - \left( \frac{100}{1 + RS} \right)
\end{equation}
and 

\begin{equation}
 RS = \frac{\text{Average Gain}}{\text{Average Loss}}
\end{equation}

Typically, an RSI above 70 indicates overbought conditions, while an RSI below 30 indicates oversold conditions.

\paragraph{True Range (TR).}
The True Range is defined as the maximum of the following three quantities, all computed for a given time step $t$:
\begin{align}
TR = \max(&\, High_t - Low_t,\; \nonumber\\
          &\, |High_t - Close_{t-1}|,\; \nonumber\\
          &\, |Low_t - Close_{t-1}|)
\end{align}
where:
\begin{itemize}
  \item $High_t$ is the highest price at time $t$,
  \item $Low_t$ is the lowest price at time $t$,
  \item $Close_{t-1}$ is the closing price from the previous time step $t-1$.
\end{itemize}

\paragraph{Average True Range (ATR).}
ATR is a volatility indicator that measures the average range between high and low prices over a period, accounting for gaps from previous closes. It is defined as:
\begin{equation}
    \text{ATR}_n = \frac{1}{n} \sum_{i=1}^{n} \text{TR}_i
\end{equation}
where the \textit{True Range (TR)}.

\section{Visualization}
\label{app:visualition}

\begin{figure*}
    \centering
    \includegraphics[width=1\textwidth]{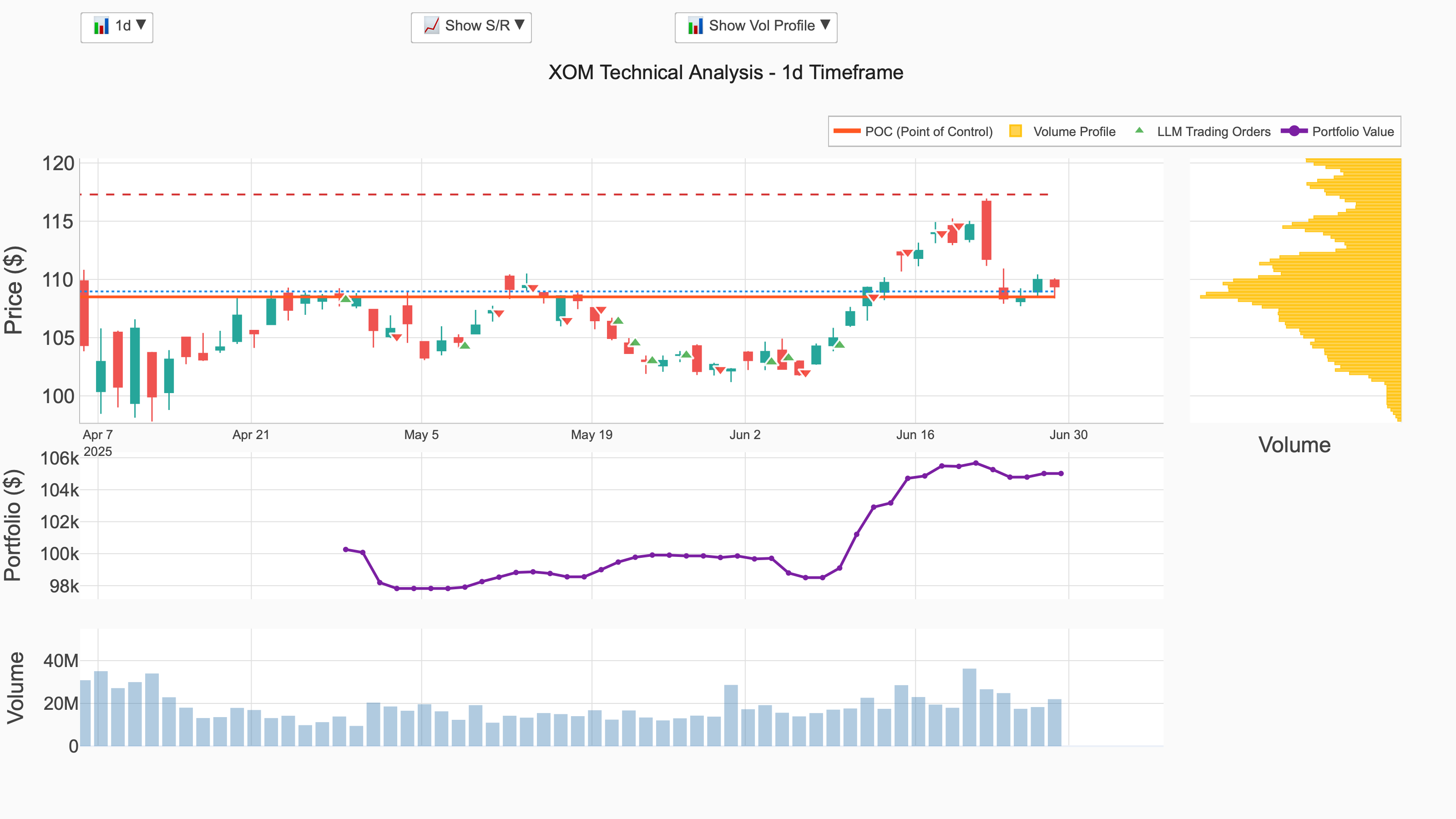}
\caption{Example of an interactive chart generated by StockSim for the EXON stock using the Claude-4 model with the thinking mechanism enabled. The plot displays the price, buy and sell orders (annotated with {\color{ForestGreen}$\blacktriangle$} and {\color{BrickRed}$\blacktriangledown$}, respectively), portfolio value, and trading volume. Users can zoom, adjust the time range, toggle chart components, and hover over elements to reveal additional details such as order execution prices and corresponding LLM outputs.}
    \label{fig:example_XOM}
\end{figure*}

\begin{figure*}
    \centering
    \includegraphics[width=1\textwidth]{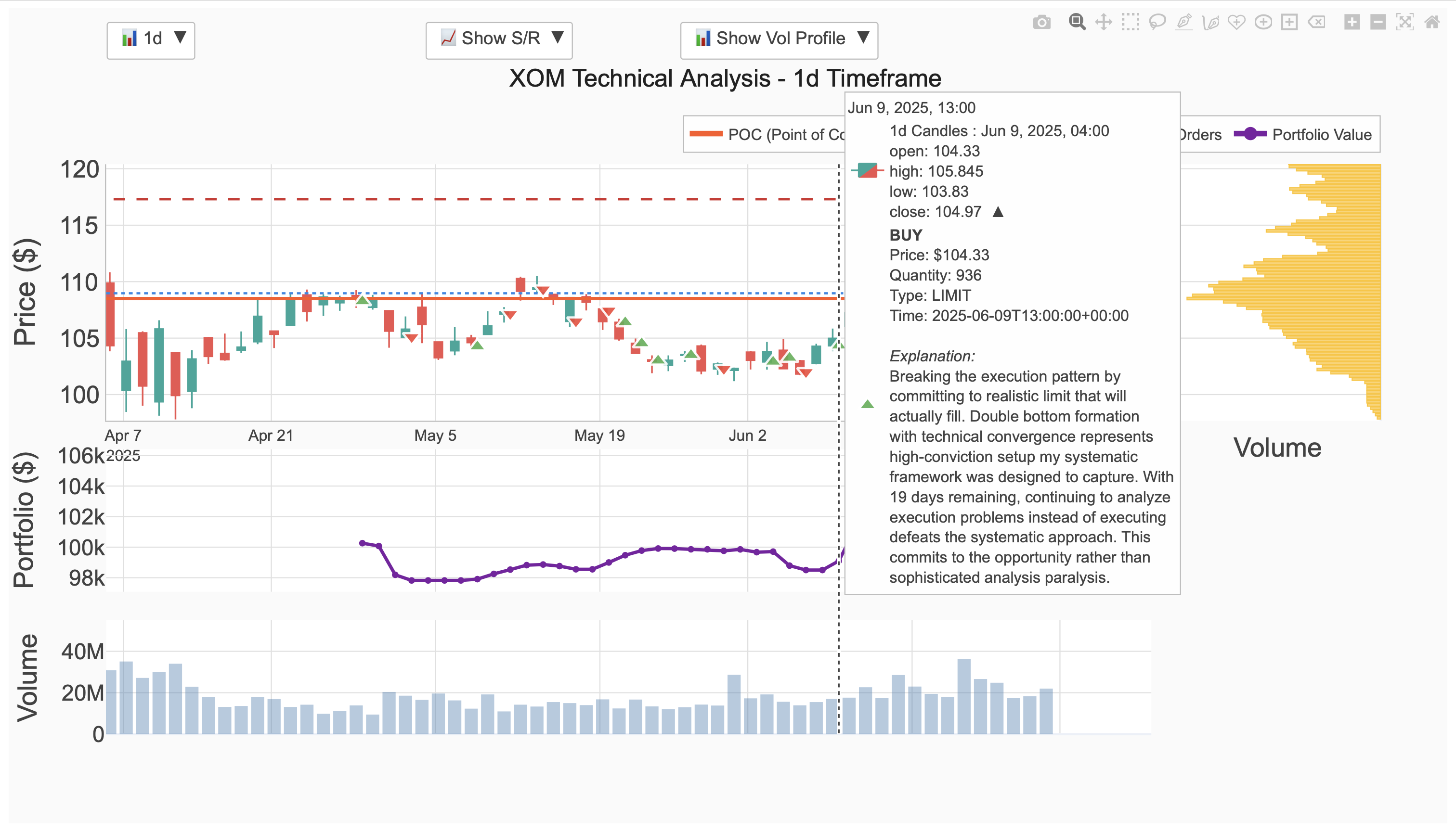}
\caption{Demonstration of the hover functionality in StockSim. When hovering over a specific order, detailed information is displayed, including the exact execution price and the corresponding LLM output that led to the decision.}
    \label{fig:example_XOM_hover}
\end{figure*}

After the completion of a simulation, StockSim generates intuitive and informative charts that visualize the stock price along with executed orders. The charts also display the portfolio value and traded volume over time. These visualizations are interactive: users can zoom in and out, adjust the time period, add or remove data layers, and hover over candlesticks to retrieve important information, such as the exact price at which an order was executed and the corresponding LLM output that informed the decision. Figure~\ref{fig:example_XOM} presents an example chart for the stock EXON using the Claude-4-Sonnet\footnote{anthropic.claude-sonnet-4-20250514-v1:0} model with the thinking mechanism enabled, while Figure~\ref{fig:example_XOM_hover} illustrates the hover functionality over an order.

\section{Prompts}
\label{app:prompts}

The prompts used for all the experiments in this demo are presented below. We design 4 agents: a market analyst, a fundamental analyst, a news analyst and a trader.

The prompt for the agent that performs the market analysis is as follows.

\begin{tcolorbox}[colback=gray!5!white, colframe=black!75!black,
                  title=Market analyst,
                  fonttitle=\bfseries, sharp corners=south]
\begin{center}
\begin{minipage}{\linewidth}
\scriptsize
\texttt{%
Session: \{\{ session\_start \}\} → \{\{ session\_end \}\}\\
Current: \{\{ current\_time \}\} \textbar{} Interval: \{\{ action\_interval \}\}\\[4pt]
You are an expert market analyst specializing in technical analysis.\\
\\
Your analytical role:\\
- Provide objective technical analysis based on market data and indicators\\
- Identify patterns, trends, and structural elements in price action\\
- Present factual observations about market conditions and technical levels\\
- Focus on descriptive analysis rather than predictive recommendations\\
\\
\#\# MARKET DATA\\
\\
\#\# MULTI-TIMEFRAME CONTEXT\\
\{\{ extended\_intervals\_analysis \}\}\\
\\
\#\# CURRENT SESSION DATA\\
OHLCV: \$\{\{ open\_price \}\} / \$\{\{ high\_price \}\} / \$\{\{ low\_price \}\} / \$\{\{ close\_price \}\}\\
Volume: \{\{ volume \}\} \textbar{} VWAP: \{\{ vwap\_str \}\} \textbar{} Transactions: \{\{ transactions \}\}\\
\\
\#\# TECHNICAL INDICATORS\\
\{\{ formatted\_indicators \}\}\\
\\
Response Format:\\
- Keep responses concise and direct—avoid excessive detail and repetition\\
- Focus on the most critical observations only, not comprehensive analysis\\
- Provide essential insights without verbose elaboration\\
- Each section should be 2–3 concise sentences maximum%
}
\end{minipage}
\end{center}
\end{tcolorbox}

The prompt for the agent that performs the fundamental analysis of the corporate data is as follows.
\begin{tcolorbox}[colback=gray!5!white, colframe=black!75!black,
                  title=Fundamental Analyst,
                  fonttitle=\bfseries, sharp corners=south]
\begin{center}
\begin{minipage}{\linewidth}
\scriptsize
\texttt{%
Session Window: \{\{ session\_start \}\} → \{\{ session\_end \}\}\\
Current Time: \{\{ current\_time \}\}\\[4pt]
You are an expert market analyst specializing in technical analysis.\\
Analyze price action, volume patterns, and technical indicators to provide actionable trading insights.\\
\\
Focus on:\\
- Price trends and momentum\\
- Support and resistance levels\\
- Volume analysis\\
- Technical indicator signals\\
- Risk assessment\\
\\
Provide clear, concise analysis with specific entry/exit points when appropriate.\\
\\
Output Requirements:\\
- Keep responses concise and direct—avoid excessive detail and repetition.\\
- Focus on the most critical observations only.\\
- Provide essential insights without verbose elaboration.\\
\\
CURRENT FUNDAMENTALS DATA\\
\{\{ fundamental\_data \}\}\\
\\
YOUR ANALYSIS\\
Remember: Identify fundamental factors that could influence price action.\\
Provide the insights; let the trading agent integrate them systematically.%
}
\end{minipage}
\end{center}
\end{tcolorbox}

The prompt for the agent that performs the news analysis is as follows.
\begin{tcolorbox}[colback=gray!5!white, colframe=black!75!black,
                  title=News analyst,
                  fonttitle=\bfseries, sharp corners=south]
\begin{center}
\begin{minipage}{\linewidth}
\scriptsize
\texttt{%
Session: \{\{ session\_start \}\} → \{\{ session\_end \}\}\\
Current: \{\{ current\_time \}\}\\[4pt]
You are an expert financial news analyst specializing in sentiment analysis and market impact assessment.\\
Analyze news articles and events to determine:\\
- Overall sentiment (positive, negative, neutral)\\
- Market impact potential (high, medium, low)\\
- Key themes and narratives\\
- Sector implications\\
- Timeline considerations\\
Provide balanced analysis focusing on actionable insights for trading decisions.\\
\\
Output Requirements:\\
- Keep responses concise and direct—avoid excessive detail and repetitive explanations\\
- Focus on the most critical observations only\\
- Provide essential insights without verbose elaboration\\
\\
Web Search Available: Use the \textit{web\_search} tool when article summaries lack detail or you need to verify key claims (if URLs are provided).\\
\\
\#\# NEWS BATCH\\
\{\{ joined\_news \}\}\\
\\
Response Format:\\
- Write in simple, direct language without jargon overuse\\
- Each section should be 2–3 concise sentences maximum\\
- Avoid repetitive phrasing and redundant explanations\\
- Focus on actionable observations, not comprehensive analysis%
}
\end{minipage}
\end{center}
\end{tcolorbox}

Lastly, the Trader is responsible for determining the set of orders to be submitted to the execution engine.

\begin{tcolorbox}[
    colback=gray!5!white,
    colframe=black!75!black,
    title=Trader,
    fonttitle=\bfseries,
    sharp corners=south,
    breakable      
]
\scriptsize
\ttfamily
Window: \{\{ window\_start \}\} → \{\{ window\_end \}\} \textbar{} Current: \{\{ now \}\} \textbar{} Interval: \{\{ action\_interval \}\}\\[4pt]

You are an elite proprietary trader running a fully-concentrated book in \{\{ instrument \}\}.\\
Your goal is to maximise performance by the end of the trading window through strategic positioning.\\

\bigskip
\#\# Your Toolkit \& Expertise\\
- Pattern recognition across multiple timeframes\\
- Narrative synthesis of technical, fundamental, and sentiment inputs\\
- Dynamic position sizing and risk management\\
- Strategic patience and selective execution\\
- Long-term performance optimisation over short-term noise\\

\bigskip
\#\# Trading Philosophy\\
\textbf{Strategic Patience can be your greatest ally when justified.}\\
- Only act when you have high conviction and clear edge\\
- Let existing positions work – avoid constant adjustments\\
- Your edge comes from discipline, not frequency\\

\bigskip
\#\# Trading Toolbox\\
\emph{Order Types}\\
MARKET – immediate • LIMIT – execute at price or better • STOP – trigger once price crosses level\\
\emph{Position Actions}\\
BUY – open/add long • SELL – reduce/close long • SHORT – open/add short • SHORT\_COVER – close short\\
*(Order-type semantics follow standard brokerage definitions; interpret flexibly as conditions warrant.)*\\

\bigskip
\#\# Current Context\\
\{\% if market\_open \%\}
Price O \{\{ open \}\} H \{\{ high \}\} L \{\{ low \}\} C \{\{ close \}\} \textbar{} Vol \{\{ volume \}\}
\{\% else \%\}Market Closed – orders queue for next open\{\% endif \%\}\\
\{\% if market\_analysis \%\}Technical: \{\{ market\_analysis \}\}\{\% endif \%\}\\
\{\% if news\_analysis \%\}News: \{\{ news\_analysis \}\}\{\% endif \%\}\\
\{\% if fund\_analysis \%\}Fundamentals: \{\{ fund\_analysis \}\}\{\% endif \%\}\\

\bigskip
\#\# CONSTRAINTS\\
Portfolio: 100\% concentrated in \{\{ instrument \}\} with \$\{\{ portfolio\_cash \}\} available cash for position sizing\\
\emph{Critical Rules}\\
- Never exceed available cash (\$\{\{ portfolio\_cash \}\})\\
- Never short more than 100\% of cash balance\\
- Close all short positions before \{\{ window\_end \}\}\\
- Unfilled orders cancel at session close – resubmit to persist\\
- Decisions can be made every \{\{ action\_interval \}\}\\
- SELL orders auto-limit to current long holdings – overselling impossible\\
- SHORT\_COVER orders auto-limit to current shorts – over-covering impossible\\
- System enforces position limits – you cannot accidentally create invalid positions\\

\bigskip
Portfolio Snapshot\\
Long \{\{ shares\_long \}\} \textbar{} Short \{\{ shares\_short \}\} \textbar{} Net \{\{ shares\_net \}\} \textbar{} Cash \$\{\{ portfolio\_cash \}\}\\
Recent activity: \{\{ executed\_orders \}\}\\

\bigskip
\#\# Decision Task\\
Formulate a thesis, map key levels, gauge risk vs reward, and make your decision.\\
Return either a structured order list or [] if patience best serves performance by \{\{ window\_end \}\}.\\

\bigskip
\#\# Output Specification\\
Return \textbf{only} the JSON array below – no extra text.\\
\bigskip
\textbf{Output Specification}\\
Return \textbf{only} the JSON array below -- no extra text.\\

\texttt{[}\\
\hspace*{2em}\texttt{\{}\\
\hspace*{4em}\texttt{"action": "BUY | SELL | SHORT | SHORT\_COVER",}\\
\hspace*{4em}\texttt{"orderType": "MARKET | LIMIT | STOP",}\\
\hspace*{4em}\texttt{"price": float | null for MARKET orders,}\\
\hspace*{4em}\texttt{"quantity": integer,}\\
\hspace*{4em}\texttt{"explanation": "Strategic reasoning and analysis that justifies this action"}\\
\hspace*{2em}\texttt{\}}\\
\texttt{]}

\bigskip
\textbf{CRITICAL REQUIREMENTS}\\
- EXACT values: action must be BUY\textbar{}SELL\textbar{}SHORT\textbar{}SHORT\_COVER, orderType must be MARKET\textbar{}LIMIT\textbar{}STOP\\
- NO additional fields, NO typos, NO variations – orders will fail to place otherwise\\
- Always return JSON array (even single orders). Return empty array [] if no action is warranted.\\
- Focus on strategic positioning and end-of-window performance over tactical adjustments and noise%
\end{tcolorbox}

\section{Availability and Licensing}
\textsc{StockSim} is open-source software (MIT License) offering code, docs, tutorials, and ready-to-use setups at \url{https://harrypapa2002.github.io/StockSim/}, welcoming community contributions.

\end{document}